\documentclass[superscriptaddress,preprintnumbers,amssymb,nofootinbib]{revtex4}
\usepackage{graphicx}
\usepackage{color}
\usepackage[latin1]{inputenc}
\usepackage{epsfig}
\usepackage{amsmath}
\usepackage{amsfonts}
\usepackage{amssymb}
\usepackage{pifont}
\usepackage{phonetic}
\usepackage{bbding}
\usepackage{eufrak}
\usepackage{bbm}

\usepackage{bbm}                                                  

\newcommand{\nc}{\newcommand}
\nc{\beq}{\begin{equation}}  \nc{\eeq}{\end{equation}}
\nc{\bea}{\begin{eqnarray}}  \nc{\eea}{\end{eqnarray}}
\nc{\baa}{\begin{array}}     \nc{\eaa}{\end{array}}
\nc{\bit}{\begin{itemize}}   \nc{\eit}{\end{itemize}}
\nc{\ben}{\begin{enumerate}} \nc{\een}{\end{enumerate}}
\nc{\bce}{\begin{center}}    \nc{\ece}{\end{center}}
\nc{\bpm}{\begin{pmatrix}}   \nc{\epm}{\end{pmatrix}}
\nc{\bvt}{\begin{verbatim}}  \nc{\evt}{\end{verbatim}}
\def\half{\frac12}	

\def\to{\rightarrow}

\def\boldoverdot{\,{\raise6pt\hbox{\bf.}\!\!\!\!\>}}

\def\lcal{{\cal L}}

\def\ocal{{\cal O}}

\def\JJ{{\bf J}}

\def\PP{{\bf P}}

\def\VV{{\bf V}}

\usepackage{bm}
\def\taubf{{\bm\tau}}		
\def\chibf{{\bm\chi}}		

\def\rhs{right hand side\ }

\def\diag{\hbox{\diag}}
\def\sm{Standard Model}

\def\gev{\hbox{GeV}}

\def\doubleundertext#1{
{\undertext{\vphantom{y}#1}}\par\nobreak\vskip-\the\baselineskip\vskip4pt%
\undertext{\hbox to 2in{}}}
\def\inbox#1{\vbox{\hrule\hbox{\vrule\kern5pt
     \vbox{\kern5pt#1\kern5pt}\kern5pt\vrule}\hrule}}
\def\sqr#1#2{{\vcenter{\hrule height.#2pt
      \hbox{\vrule width.#2pt height#1pt \kern#1pt
         \vrule width.#2pt}
      \hrule height.#2pt}}}

\def\today{\ifcase\month\or
  January\or February\or March\or April\or May\or June\or
  July\or August\or September\or October\or November\or December\fi
  \space\number\day, \number\year}
\def\pmb#1{\setbox0=\hbox{#1}%
  \kern-.025em\copy0\kern-\wd0
  \kern.05em\copy0\kern-\wd0
  \kern-.025em\raise.0433em\box0 }

\def\pmbb#1{\setbox0=\hbox{#1}%
  \kern-.02em\copy0\kern-\wd0
  \kern.04em\copy0\kern-\wd0
  \kern-.02em\raise.03464em\box0 }
\def\up#1{^{\left( #1 \right) }}
\def\inv#1{\frac1{#1}}
\def\su#1{{SU(#1)}}

\def\sumprime_#1{\setbox0=\hbox{$\scriptstyle{#1}$}
  \setbox2=\hbox{$\displaystyle{\sum}$}
  \setbox4=\hbox{${}'\mathsurround=0pt$}
  \dimen0=.5\wd0 \advance\dimen0 by-.5\wd2
  \ifdim\dimen0>0pt
  \ifdim\dimen0>\wd4 \kern\wd4 \else\kern\dimen0\fi\fi
\mathop{{\sum}'}_{\kern-\wd4 #1}}
\def\lsim{\:\raisebox{-0.5ex}{$\stackrel{\textstyle<}{\sim}$}\:}
\def\gsim{\:\raisebox{-0.5ex}{$\stackrel{\textstyle>}{\sim}$}\:}
\def\leff{\lcal_{\rm eff}}
\def\dm{\delta m_h^2}

\def\ti{{\rm SM}}
\def\tii{{\rm Hvy}}
\def\tiii{{\rm eff}}

\begin{document}

\title{EFT-naturalness: an effective field theory analysis of Higgs naturalness}
\author{Shaouly Bar-Shalom}
\email{shaouly@physics.technion.ac.il}
\affiliation{Physics Department, Technion-Institute of Technology, Haifa 32000, Israel}
\author{Amarjit Soni}
\email{adlersoni@gmail.com}
\affiliation{Theory Group, Brookhaven National Laboratory, Upton, NY 11973, USA}
\author{Jose Wudka}
\email{jose.wudka@ucr.edu}
\affiliation{Physics Department, University of California, Riverside, CA 92521, USA}
\date{\today}

\begin{abstract}
Assuming the presence of physics beyond the \sm\ (SM) with a characteristic scale
$M \sim {\cal O}(10)$ TeV, we investigate the naturalness of the Higgs sector at scales below $M$
using an effective field theory (EFT) approach. We obtain the leading 1-loop EFT contributions to the Higgs mass with a Wilsonian-like hard cutoff, and determine
the constraints on the corresponding operator coefficients for these effects to alleviate the
{\it little hierarchy problem}
up to the scale of the effective action $ \Lambda < M $, a condition we denote
by ``EFT-naturalness''.
We also determine the types of physics that can lead to EFT-naturalness and
show that these types of new physics are best probed in vector-boson and multiple-Higgs production.
The current experimental constraints on these coefficients
are also discussed.
\end{abstract}


\maketitle

\bigskip\paragraph{Introduction} $~$\\
\\
The recent LHC discovery of a light 126 GeV scalar particle
\cite{LHCHiggs}
brought us one step closer to
understanding the mechanism of electroweak symmetry breaking.
Indeed, the measurements of its
production and decays to the SM's gauge-bosons \cite{PDGHiggs}
are consistent (within large errors) with the SM.
Moreover, in view of the fact that no evidence for new physics has been
observed yet up to energies of $\sim 1-2$ TeV and that
the SM with a 126 GeV Higgs seems to be a
consistent theory up to the Planck scale
(favoring a metastabe EW vacuum \cite{PDGHiggs,lambda1,lambda2}),
this discovery exacerbates the long-standing fundamental difficulty of the SM
known as the hierarchy problem. Simply put, the presence of a fundamental
Higgs with an EW-scale mass appears unnatural, since if the SM is the only physics
present up to some high scale $\Lambda$, it is then hard
to see why the Higgs boson mass $m_h $ does not receive large corrections of
$O(\Lambda)$. This technical difficulty is also known as the naturalness or
fine-tuning problem of the SM. It becomes evident when one calculates the SM's leading
$O(\Lambda^2)$ 1-loop
corrections to the Higgs mass squared with a hard cutoff:
\beq
\dm(\ti) = \frac{\Lambda^2}{16 \pi^2}
\left[ 24 x_t^2 -6 \left(2 x_W^2 + x_Z^2  + x_h^2 \right) \right]
\sim 8.2 \frac{\Lambda^2}{16 \pi^2} ~~, ~~x_i \equiv
\frac{m_i}{v} ~~(v\simeq246 \gev)\label{deltaSM} ~,
\eeq
where the dominant contribution is
generated by the top-quark loop.
This gives
$\dm(\ti) \approx m_h^2$
already for $\Lambda \sim 550$ GeV when $ m_h \sim 125$ GeV and the Higgs mass is then said to
be unnatural above this scale.
In Wilson's approach \cite{wilson}, the hard cutoff $\Lambda$ in Eq.~\ref{deltaSM} corresponds
to the scale of the effective action - in the following we will use this picture to 
investigate the behavior of $\dm$ in the presence of new physics (NP) with a mass scale $M > \Lambda$.

In particular, if we now imagine the presence of NP with
a characteristic scale $M > \Lambda > v$, then Eq.~\ref{deltaSM}  will be modified as the  heavy excitations will
generate new contributions to $\dm $. These contributions can be derived within specific models (e.g., little-Higgs or supersymmetric theories,
or phenomenological extensions of the SM with additional heavy scalars and/or fermions \cite{jose-pragmatic,jose-DM-singlets,craig}), or one can adopt a
model-independent approach using an effective field theory (EFT).
The first approach assumes full knowledge of the physics up to a yet higher scale
(i.e., larger than $M$), above which the selected model breaks down
(or is subsumed by a more fundamental theory); accordingly, in this case the scale
of the effective action, $ \Lambda $, can be extended
beyond $M$, i.e., beyond the typical mass scale of the new particles of a
specific theory.

In contrast, the EFT approach refrains from selecting a specific model but allows a reliable calculation of $ \dm $  only when the cutoff is below $M$;  this
approach can be used to impose general restrictions on the parameters of the unknown
underlying theory by imposing the condition that the EFT remains natural for all scales $\Lambda < M$.

In this paper we wish to investigate naturalness using this EFT approach,
following Wilson's prescription with a hard cutoff and
assuming also that {\it the NP is weakly coupled and renormalizable}
(or, alternatively, that all non-renormalizable terms in the theory
are suppressed by inverse powers of a
much higher scale $\gg M $).
Thus, at scales below $M$ the NP is not directly observable\footnote{At scales above $M$
the NP becomes manifest, and
naturalness issues related to quadratic divergences may arise
in connection with new heavy scalar particles
that might be present, but such complications will not affect the conditions under which heavy
new physics may tame the little hierarchy problem at scales {\em below} $ \Lambda $, which is our
only concern in this paper. Note that fermionic solution(s) do not suffer from this difficulty.}
but it can have important virtual effects that generate both
renormalization of the SM parameters and  an infinite tower of
effective operators with dimension $ \ge 5 $.
We will then  evaluate the effects of these higher dimensional operators
to obtain ``EFT-naturalness".  Namely, the conditions
and relations among the EFT parameters
for which naturalness in the SM Higgs sector can be ameliorated, i.e.,
addressing the little hierarchy problem of the SM Higgs sector
up to the scale of the effective action $\Lambda$.
We note in passing that higher dimensional NP operators in the SM Higgs
sector may also have a significant effect on the stability of the EW vacuum \cite{stability-NP}.

We denote such higher dimensional operators by
$\ocal_i\up n$ ($n$ denotes the dimension and $i$  all other
distinguishing labels), which are local, gauge and Lorentz
invariant combinations of SM fields and their derivatives.
They result from integrating out the heavy
degrees of freedom  of the heavy NP theory
that underlies the SM, and expanding in inverse powers
of $M$ after appropriate renormalization of the SM parameters.\footnote{We adopt
the minimal coupling scheme in constructing
the higher dimensional effective operators below, which is consistent
with the assumption of a weakly coupled and renormalizable underlying NP,
see e.g., \cite{trott1};
the compatibility of minimal coupling with EFT was recently discussed
in the literature (see \cite{trott1,spanish}).}

The effective Lagrangian then takes the form \cite{buchwyler,patterns,Wudka-new}:
\begin{eqnarray}
\leff = \sum_{n=5}^\infty \frac1{M^{(n-4)}}  {\sum_i
f_i^{(n)}  \ocal_i^{(n)}}\label{eff1} ~.
\end{eqnarray}
Different types of
NP can generate  the {\em same} operators, but, in general,
with {\em different} coefficients, so that the SM renormalization
constants and the operator coefficients
parameterize all possible types of NP. Moreover, some of the $ \ocal $'s
are {\em necessarily} generated by loops involving only
heavy particles \cite{patterns,Wudka-new}
and we label such operators `loop-generated' (LG); this
is a useful distinction because graphs involving
LG operators and $\ell$ SM loops are considered to be at least $\ell+1$ loop diagrams.

\begin{figure}[htb]
\includegraphics[scale=0.5]{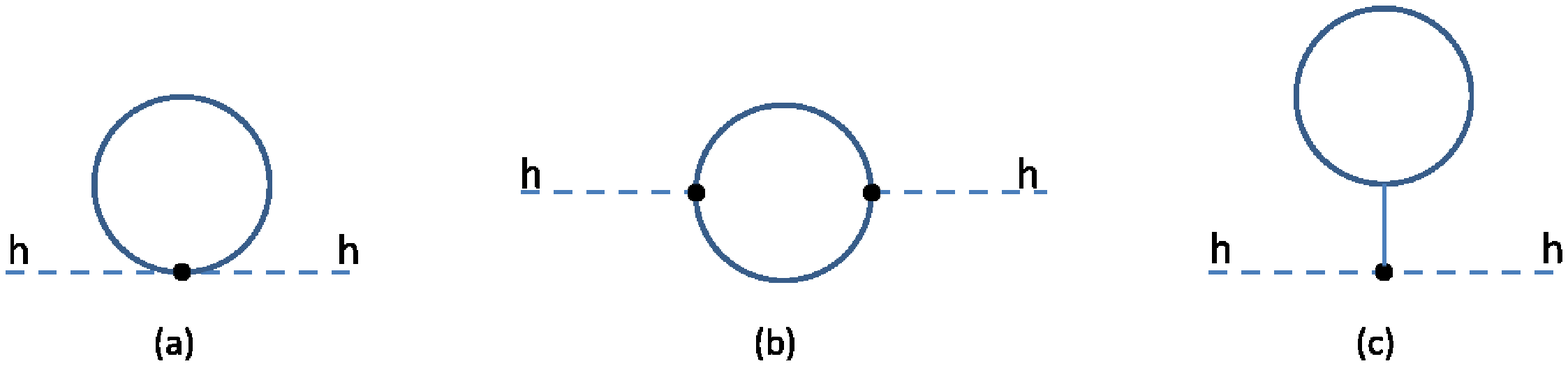}
\caption{\emph{The 1-loop graphs generating $ \dm $. The internal
lines represent bosons or fermions from either the SM or the heavy NP.}
\label{fig:graphs}}
\end{figure}

\bigskip\paragraph{EFT and the one-loop Higgs mass corrections} $~$\\
\\
In general, {\em all} (SM and NP) one-loop corrections to $ m_h^2 $ are generated by the graphs in
Fig.~\ref{fig:graphs}. In the scenarios we are interested in here,
these corrections can be separated into 3 categories:
\begin{description}
\item{\boldmath{$\dm(\ti)$}:} When all internal lines are the light SM fields.
The contributions from this category
are given in Eq.~\ref{deltaSM}.
\item{\boldmath{$\dm(\tii)$}:} When all internal lines are heavy fields of the underlying NP.
The contributions
from this category are
contained in the renormalization of the parameters of the SM
that follows upon integration of the heavy particles.
This is included in what we denote here as
``tree-level'' parameters, i.e.,
$m_h^2({\rm tree}) = m_h^2({\rm bare}) + \dm(\tii)$.
\item{\boldmath{$\dm(\tiii)$}:} When one line is heavy and the other
is light (in graphs (b) and (c) in Fig.~\ref{fig:graphs}).
The contributions in this category are generated by the
effective Lagrangian in Eq.~\ref{eff1}
and are the ones we are interested in here.
\end{description}

As noted earlier, the little
hierarchy problem of the SM refers to the fact
that $ \dm(\ti) > m_h^2({\rm tree}) $ when $ \Lambda \gtrsim 500~\gev $,
assuming $m_h ({\rm tree}) $ is close to the observed value
$m_h({\rm tree}) \simeq m_h\simeq 125~\gev $.
Our aim here is only to address this problem at scales below $\Lambda$, viewed as the scale of 
the Wilsonian effective action; we will not be concerned with the issues related
to the UV-completion of the SM or EW-Planck hierarchy, or with
any details of the underlying theory giving rise to Eq.~\ref{eff1}.
Specifically, we will
study the role that the effective interactions in Eq.~\ref{eff1} may play in
restoring naturalness to the Higgs sector at any given intermediate scale 
$v < \Lambda < M$,\footnote{It is important to note that Eq.~\ref{eff1} can be used to
calculate such NP effects provided all energies (including those that appear within loop
calculations) are kept below $ M $ so the cutoff must obey $ \Lambda < M$.}
and determine
the conditions under which
$ \dm(\ti) + \dm(\tiii) \lesssim m_h^2 $ when $ m_h \ll \Lambda \le M$.
We now proceed to the calculation of $ \dm(\tiii)$.

\begin{figure}[htb]
\includegraphics[scale=0.5]{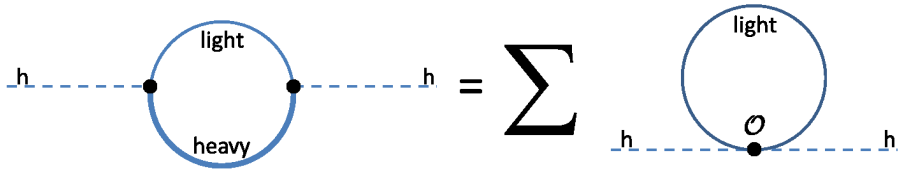}
\caption{\emph{Description of the manner in which the effective
Lagrangian in Eq.~\ref{eff1} generates graphs in category $\dm(\tiii)$
defined in the text. } \label{fig:expansion}}
\end{figure}

To illustrate the manner in which the effective operators contribute to $\dm$ we consider the contributions
to the Higgs mass generated by the diagrams in Fig. \ref{fig:expansion}.
Expanding the heavy propagator in powers of its (large) mass $M$,
one generates an infinite series of vertices suppressed by inverse powers of $M$
(see  Fig.~\ref{fig:expansion} for a schematic depiction). As mentioned above, we will evaluate
loop graphs using a cutoff prescription (with $ \Lambda$ being the cutoff) so that this expansion
remains valid for the graphs in Fig.~\ref{fig:expansion} (recall that $ M \ge \Lambda $)
and the effective vertices correspond
to those  generated by the effective operators in Eq.~\ref{eff1}.
We therefore need the set of operators
$ \ocal $ which are   not LG, and contribute to graph
Fig.~\ref{fig:graphs}(a), where the vertex is generated by the effective
operator.

The non-LG operators that give $O(\Lambda^2) $
contributions to $ \dm $ can be characterized using the
following arguments: the internal lines in the graphs
on the \rhs of  Fig.~\ref{fig:expansion}
(with the $ \ocal$-generated vertices) can be either the
SM scalar, fermions or vectors. For the first case,
$ \ocal $ must contain at least 4 SM scalar doublets;
but if it contains more than 4 such scalar fields
the corresponding contributions to $ \dm $ are
suppressed by powers of $ (v/M) $ and are, therefore,
subdominant.
Thus, leading contributions with a
scalar internal line are generated by effective operators with
precisely 4 scalar doublets.
Similarly, if the internal lines are fermions or vectors,
the operators must contain 2 SM scalar doublets.
Lastly, it is straightforward to show \cite{patterns,Wudka-new}
that operators with $2$ SM scalar doublets, no fermions
and any number of vectors are LG and are also subdominant.

Summarizing:
the operators that generate 1 loop $O(\Lambda^2) $ contributions to $ \dm$
can be of {\it only} two types:
\bit
\item Type I: $ \ocal $  contains 4 scalar fields, any number of derivatives and is not LG.
\item Type II: $ \ocal $  contains 2 fermions and 2 scalar fields, any number of derivatives and is not LG.
\eit

\begin{figure}[htb]
\includegraphics[scale=0.45]{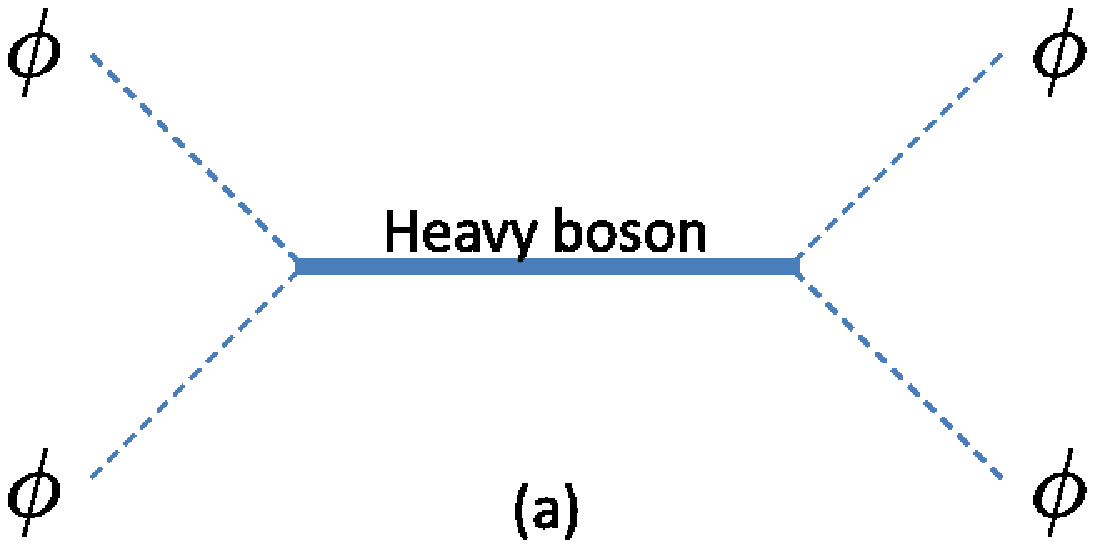} \quad
\includegraphics[scale=0.45]{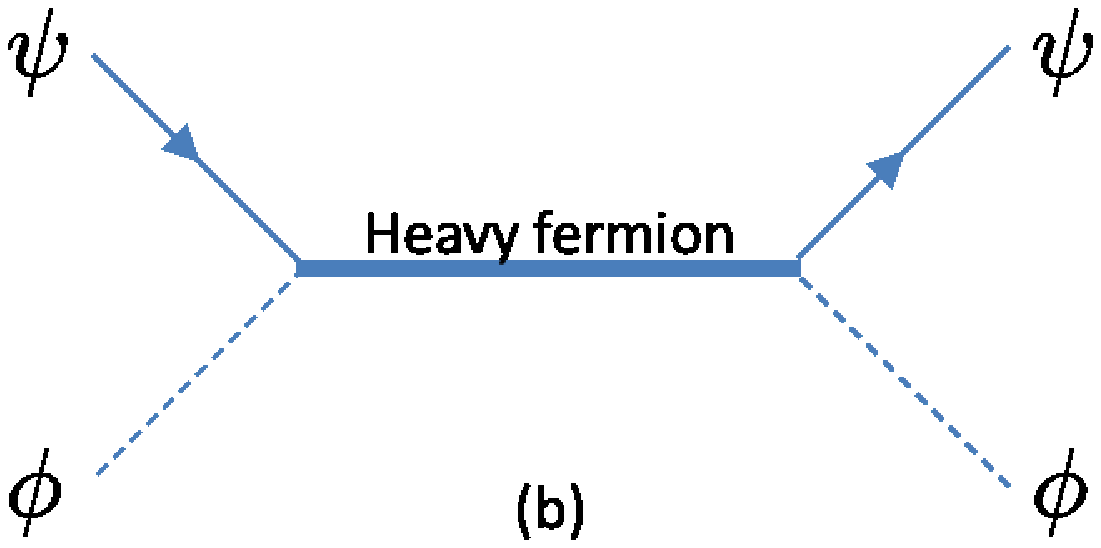} \quad
\includegraphics[scale=0.45]{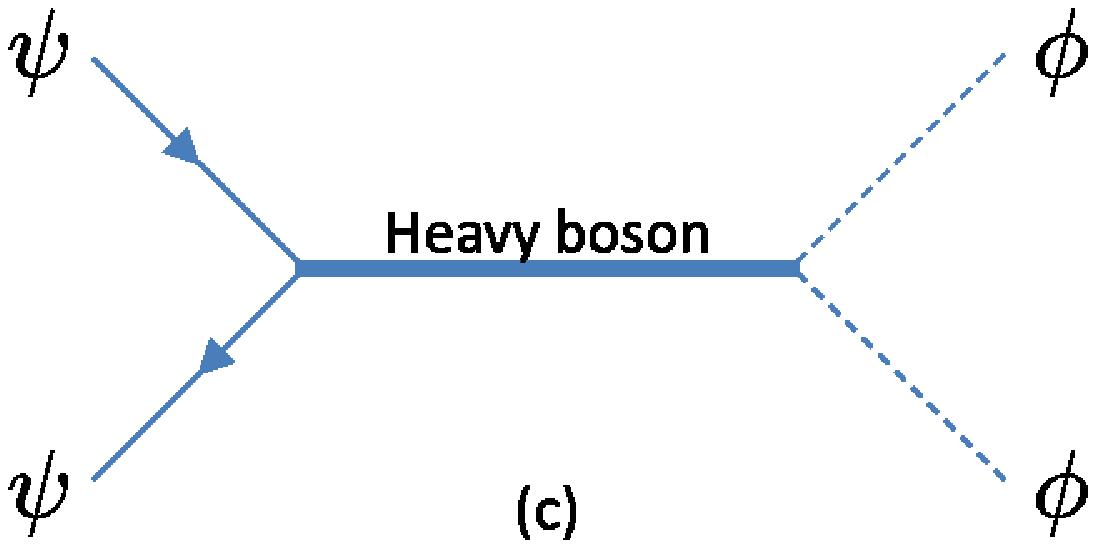}
\caption{\emph{Tree-level graphs that generate the effective operators
of type I (diagram a) and II (diagrams b and c),
that can produce leading corrections to $ \dm $. $ \phi $ and $ \psi $ denote
the SM scalar doublet and fermions, respectively and all vertices
are understood to be invariant under SM gauge transformations.}
\label{fig:np}}
\end{figure}

The simplest way to determine the form of the operators of types I and II
is by recalling that these operators are generated
at tree-level in the underlying heavy theory by the graphs in Fig.~\ref{fig:np},
where the relevant $\ocal$ is obtained by expanding the
propagators in inverse power of the internal heavy mass and imposing gauge invariance.
We can also eliminate operators with derivatives that
act on scalar fields connected to the external legs,
since $ \dm $ is evaluated at zero momentum.\footnote{If
$ \dm $ is evaluated at some other low scale, e.g. at $ \mu = m_h $,
then vertices where a derivative acts on an external leg are also
subdominant since their contribution will be suppressed by a
factor of $ m_h/\Lambda $.}

A further simplification follows from a more
careful study of the diagram shown in Fig.~\ref{fig:np}(c), which
contributes to $ \dm $ only through graphs of the type depicted in
Fig.~\ref{fig:graphs}(c), for which the heavy boson must be
a scalar. This heavy scalar must also be an $\su2$ triplet
or singlet (since it couples to two SM isodoublets), which implies
that the fermions must have the same chirality (since a pair of fermions
with different chirality cannot form a singlet or a triplet).
It then follows that
the loop in Fig.~\ref{fig:graphs}(c) must involve a chirality
flip, so that its contribution to $\dm$ will
be suppressed by a factor of $ m_\psi/\Lambda $
and is, therefore, also subdominant.
We thus conclude that we can neglect the effective
operators associated with Fig.~\ref{fig:np}(c).

With the above comments it is a straightforward exercise to obtain the
relevant set of effective operators of interest.
Those generated by heavy scalar exchanges in Fig.~\ref{fig:np}(a) are
\beq
\ocal_S\up{2k+4} = \half |\phi|^2 \Box^k |\phi|^2~, \quad
\ocal_{\chibf}\up{2k+4} = \half (\phi^\dagger \tau_I \phi) D^{2k}
(\phi^\dagger \tau_I \phi)~, \quad
\ocal_{\tilde\chibf}\up{2k+4} = \inv4 (\phi^\dagger \tau_I \tilde\phi) D^{2k}
(\tilde\phi^\dagger \tau_I \phi)
\label{eq:s-ops} ~,
\eeq
which correspond to the cases where the heavy scalar is a SM gauge singlet (labeled $S$)
or an isotriplet of hypercharge 0 or 1 (labeled $ \chibf$ and $ \tilde \chibf$, respectively).
There are no other operators of this type since $S,~\chibf$ and $ \tilde \chibf$ are the only
possible three states
that can be formed with two SM scalar isodoublets. In the following we denote these
heavy scalars collectively by  $ \Phi $.

Similarly the operators generated by heavy vector exchanges in Fig.~\ref{fig:np}(a) are
\beq
\ocal_v\up{2k+6} = \half j _\mu \Box^k j ^\mu ~, \quad
\ocal_{\tilde v}\up{2k+6} =  \tilde j^\dagger_\mu \Box^k \tilde j^\mu ~, \quad
\ocal_{\VV}\up{2k+6} = \inv6 J_{I \mu} D^{2k} J^\mu_I
\label{eq:v-ops}~,
\eeq
where the currents are
\beq
j^\mu = i \phi^\dagger  D^\mu \phi + {\rm H.c.} ~,\quad
\tilde j^\mu = i \tilde\phi^\dagger  D^\mu \phi ~,\quad
 J^\mu_I =i \phi^\dagger \tau^I D^\mu \phi + {\rm H.c.}
\label{eq:j}~,
\eeq
and the labels in Eq.~\ref{eq:v-ops} refer to heavy vector isosinglets ($v,\,\tilde v$)
of hypercharge 0 or 1, respectively,
and a heavy vector isotriplet ($\VV$) of hypercharge 0.
In the following we will collectively denote these heavy
vectors by $X$.\footnote{There is a fourth current that can be constructed
using 2 scalar fields, namely
$\tilde \JJ^\mu =i \tilde\phi^\dagger \taubf D^\mu \phi$; however, since
$ \tilde \JJ^\mu = D^\mu \PP $ with $ \PP \equiv (i/2) \tilde\phi^\dagger \taubf \phi$
holds identically, and since vector bosons do not have tree-level couplings
to total derivatives, there are no tree-level operators involving $ \tilde J$.
None of the other currents in Eq.~\ref{eq:j} can be written as derivatives of scalar operators.}

Finally, the graph in Fig.~\ref{fig:np}(b) involves
 an exchange of a heavy fermion $ \Psi $ which may or may not be colored
 and has the same quantum numbers as
$ \phi \psi $ or $ \tilde \phi \psi $. That is, $\Psi$ can be
an isosinglet, doublet
or triplet heavy lepton or quark of hypercharge $ y_\Psi = y_\psi \pm1/2 $ ($y_r$ denotes
the hypercharge of $r$). These $\Psi$-generated operators  are
\beq
\ocal\up{2k+4}_{\Psi -\psi}=
|\phi|^2\,  \bar\psi \left(i \not\!\!D\right)^{2k-1} \psi, \quad (k\ge1)
\label{eq:f-ops}~,
\eeq
where $ \psi $ is any SM fermion.\footnote{It
is, in principle, possible to eliminate the operator in Eq.~\ref{eq:f-ops}
using the ``equivalence theorem" of \cite{Wudka-new}.
However in a cut-off scenario like the one we consider, this procedure
involves a non-trivial Jacobian that will generate terms of the form
$ |\phi|^2 \Lambda^2 $, which will reproduce the contributions
to $ \dm $ generated by $\ocal_{\Psi-\psi} $.}
Another type of operator that may be generated by the heavy-fermion exchange is
$( \phi^\dagger \tau_I \phi ) \, (\bar\psi \tau_I\not\!\!D^{2k-1} \psi)$,
where $ \psi $ is an isodoublet. However, this operator
will yield a contribution to $ \dm$ which is
suppressed by a factor of $ m_\psi^2/\Lambda^2 $ and is, therefore, also subdominant.

Note that the graphs in Fig.~\ref{fig:np} represent the possible types of NP that
can generate the effective operators in Eqs.~\ref{eq:s-ops}, \ref{eq:v-ops} and \ref{eq:f-ops} at tree-level. There are other
types of NP that can also generate these operators, but only via loop
diagrams. It then follows that the coefficients of the
operators associated with the same heavy particle are correlated; we
return to this point below.

Calculating the 1-loop quadratic corrections to $m_h^2$ which are generated by the operators in
Eq.~\ref{eq:s-ops}, \ref{eq:v-ops} and \ref{eq:f-ops}, we obtain:
\beq
\dm(\tiii) = -\frac{\Lambda^2}{16 \pi^2} F\up{\rm eff} ~,
\label{deltaeff}
\eeq
where ($\Phi= S,~\chibf,~\tilde \chibf$ and $ X = v,~\tilde v,~\VV$)\footnote{When
evaluating the contributions associated with the heavy vectors (sum over $X$)
it is convenient to use a renormalizable gauge. The calculation in the unitary gauge
is more involved since, in this case, the longitudinal component of the vector
propagators are proportional to $1/v^2$, and this complicates isolating the leading
contributions.}
\beq
F\up{\rm eff}=
\sum_{k=0}^\infty  \frac{(\Lambda/M)^{2k}}{k+1}  \sum_\Phi f_\Phi\up{2k+4} -
\sum_{k=0}^\infty  \frac{(\Lambda/M)^{2k+2}}{k+2}  \sum_X f_X\up{2k+6} -
\sum_{k=1}^\infty \frac{(-1)^k(\Lambda/M)^{2k}}{k+1}  \sum_{\Psi,\psi} f_{\Psi-\psi}\up{2k+4}
 ~. \label{dmh.eff}
\eeq

Defining the measure for fine-tuning to be $\Delta_h \equiv |\dm|/m_h^2$,
where $\dm =  \dm(\ti) + \dm(\tiii) $
and $m_h^2$ is the physical mass, $ m_h^2 = m_h^2({\rm tree}) + \dm $,
we have:\footnote{Our measure for naturalness corresponds to what is known as
the Barbieri-Giudice criteria \cite{BGcriteria}:
$\Delta = |\partial ln O|/|\partial ln f|$, for
$O = m_h^2$ and $f = F\up{\rm eff} - 8.2$.}
\begin{eqnarray}
\Delta_h = \frac{\Lambda^2}{16 \pi^2 m_h^2} \left| F\up{\rm eff} - 8.2 \right|
\label{ft}~.
\end{eqnarray}

\begin{figure}[htb]
\includegraphics[scale=0.8]{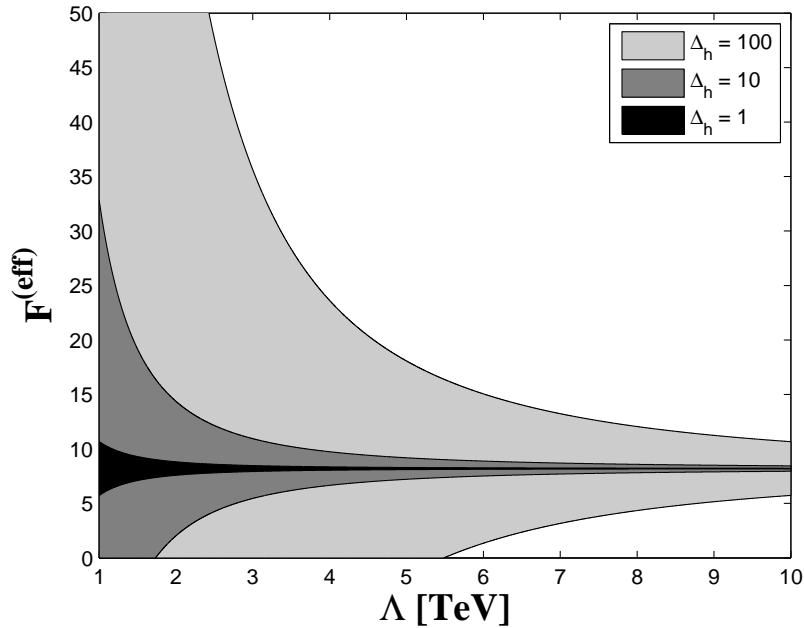}
\vspace{-7.5cm}
\caption{\emph{Regions in the $F\up{\rm eff}-\Lambda$ plane
where naturalness can be restored with no fine-tuning
($\Delta_h = \dm/m_h^2 = 1$, in black) and with fine-tuning at the level
of 10\% (dark gray) and 1\% (light gray), corresponding to
$\Delta_h = \dm/m_h^2 =10$ and 100, respectively. See also text.}
\label{fig3}}
\end{figure}

In order to restore naturalness at 
the effective action scale $\Lambda < M$, there must be
a cancellation\footnote{If no cancellation occurs between the radiative corrections then $ \dm = O(\Lambda^2) $, which must be
balanced by an $O(\Lambda^2)$ tree-level contribution to the Higgs mass in order to explain
the experimentally observed value of the physical mass $ m_h^2 = O(v^2) $.} between the
$O(\Lambda^2)$ one-loop expressions generated in the SM (Eq.~\ref{deltaSM}) and those produced by
the effective operators (Eq.~\ref{deltaeff}); this cancellation can be  partial or exact (e.g., due to
a model/symmetry), leading to $ \dm = O(v^2 ) $ -- in this case the tree-level contribution $m_h^2(\rm tree)$ will be of this order as well.
It should be emphasized that $F\up{\rm eff}$ is a dimensionless function of the
NP parameters and the ratio $ \Lambda/M$, so that the cancellation conditions will depend on $ \Lambda$ as well. Our requirement that $ \Lambda $ be the scale below which the SM little-hierarchy problem is solved is consistent within our EFT-naturalness scenario because of the requirement $ M>\Lambda $.

Re-writing the above defined fine-tuning condition
as $ | m_h^2({\rm tree})/\dm +1|
=1/\Delta_h$, it is evident that this cancellation  must occur to a precision of
$ 1/\Delta_h $, so that a larger $\Delta_h$ corresponds to a less natural
theory. Inspection of Eq.~\ref{ft} shows that large values of $ \Delta_h$ also
correspond to a less stringent correlation between $\dm(\ti)$ and $\dm(\tiii)$:
this naturality criterion refers to the relationship between tree
and radiative corrections, {\em not} to relationships among the various
radiative corrections.
Therefore, a theory (i.e., $F\up{\rm eff}$) for which $\Delta_h=1$ is natural,
while one with $\Delta_h=10(100)$ suffers from fine-tuning of  10\%(1\%).

In Fig.~\ref{fig3} we plot regions in the $F\up{\rm eff} - \Lambda$ plane
that correspond to
an effective action which is natural (i.e., enclosed within the $\Delta_h = 1$ region) and
those that suffer from fine-tuning of no worse than 10\% and 1\%,
corresponding to $ \Delta_h=10$ and $\Delta_h=100$, respectively.
For example,
theories for which  $8.17 \lesssim F\up{\rm eff}  \lesssim 8.23$
are natural at $\Lambda \sim 10$ TeV, while theories with
$7.95 \lesssim F\up{\rm eff}  \lesssim 8.45$ or
$5.73 \lesssim F\up{\rm eff}  \lesssim 10.67$ will suffer from
10\% or 1\% fine-tuning, respectively, at $\Lambda \sim 10$ TeV.
Note also that, if the scale of EFT-naturalness
is $\Lambda \sim 5$ TeV, then a much wider range of theories,
those giving $0 \lsim F\up{\rm eff}  \lsim 18$, are allowed
if one is willing to tolerate 1\% fine-tuning.
It should also be noted that the EFT-naturalness regions
shown in Fig.~\ref{fig3} may in general be subject to additional constraints
(e.g., from perturbativity),
depending on the details of the specific underlying theory.
\begin{figure}[htb]
\includegraphics[scale=0.55]{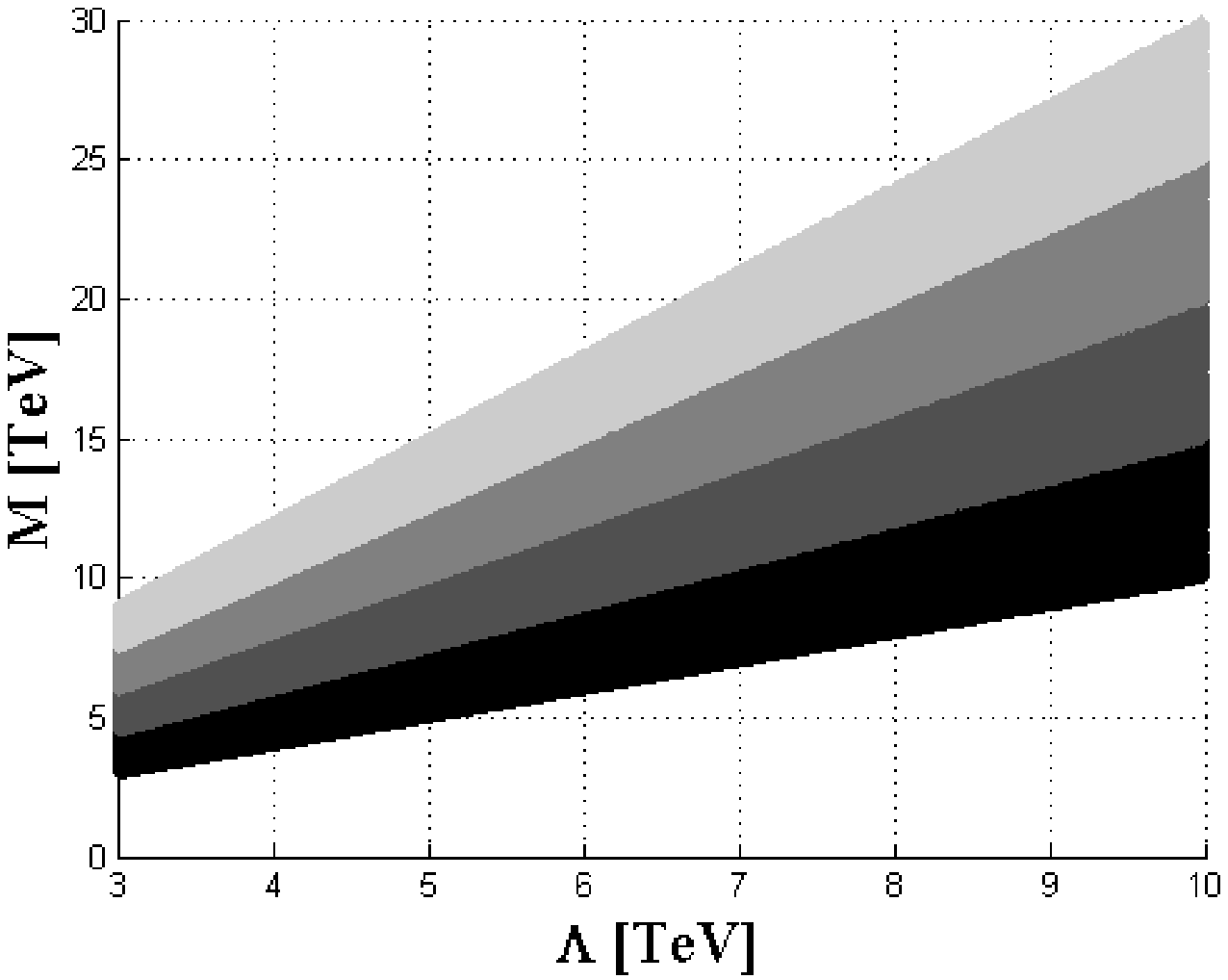} \\
\includegraphics[scale=0.55]{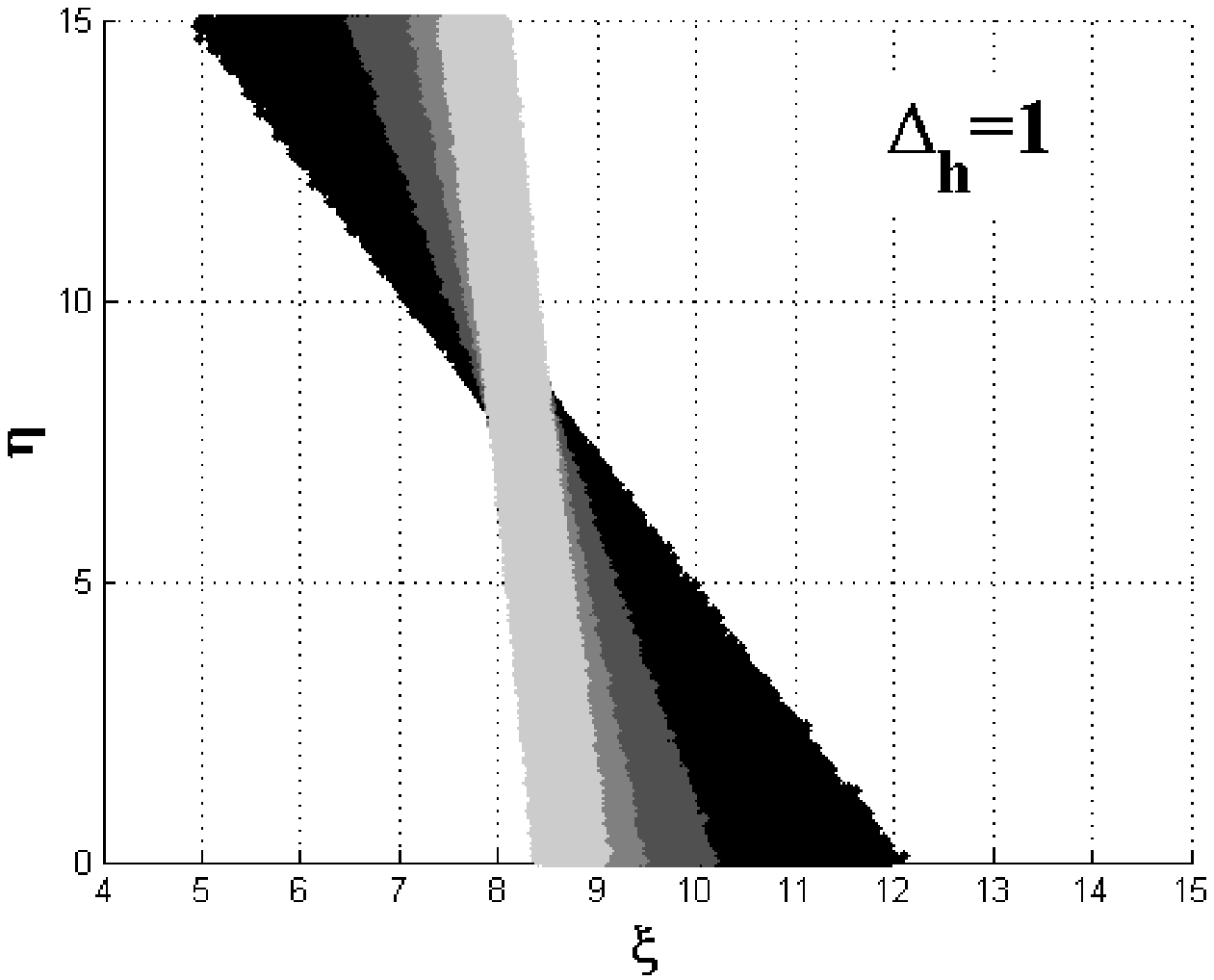}
\includegraphics[scale=0.55]{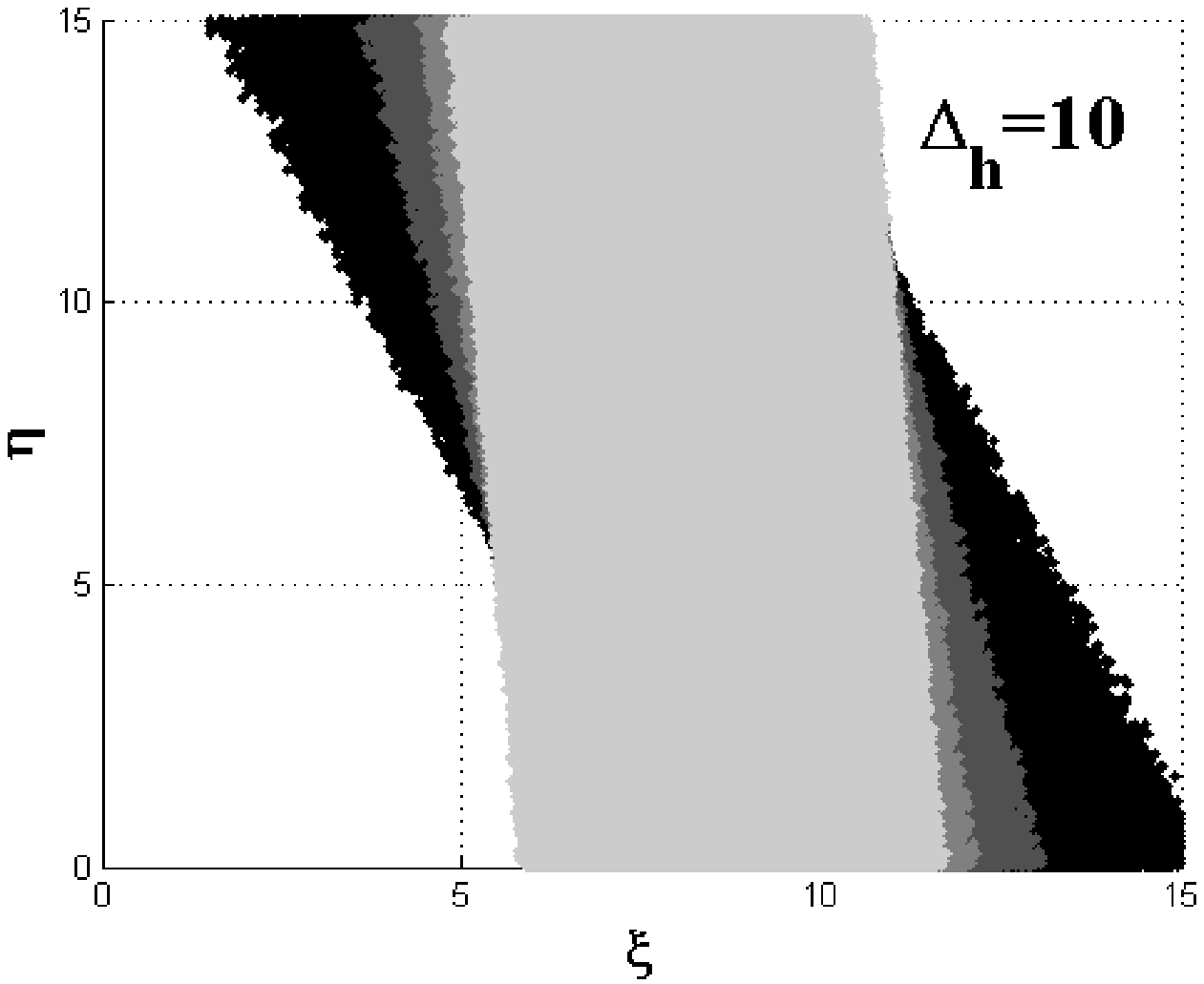}
\caption{\emph{Upper plot: regions in the $\Lambda - M$ plane, corresponding
to $1< M/\Lambda <1.5$ (black), $ 1.5< M/\Lambda <2$ (dark gray),
$ 2< M/\Lambda <2.5$ (gray) and $ 2.5< M/\Lambda <3$ (light gray), for
a naturalness scale $3~{\rm TeV} < \Lambda < 10 ~{\rm TeV}$ and $M$ being the
typical NP mass scale. Lower plots:
scatter plots in the $\xi - \eta$ plane (see Eq.~\ref{xietaeq})
corresponding to the regions in the  $\Lambda - M$ plane (corresponding shading colors), where
the NP (with scale $M$) restores naturalness (left plot) or suffers from 10\%
fine-tuning (right plot).}
\label{xieta}}
\end{figure}

Given the specific form of the graphs in Fig.~\ref{fig:np} which generate the leading operators
in Eqs.~\ref{eq:s-ops}, \ref{eq:v-ops} and \ref{eq:f-ops}, it is possible to express the EFT coefficients
$f$ in terms of some of the couplings of the heavy particles to the SM. Specifically,
defining $u_\Phi$, $g_X$ and $y_{\Psi-\psi}$ to be the couplings
of a heavy scalar
$\Phi = S,\,\chibf,\,\tilde\chibf$ to $\phi^2$ (i.e., $u_\Phi \phi^\dagger\Phi \phi $), of a heavy vector
boson $ X= v,\,\tilde v,\,\VV$ to the currents $J_X = j,\,\tilde j,\,J_I$ in Eq.~\ref{eq:j} (i.e., $g_X X_\mu J_X^\mu$),
and of a heavy fermion $\Psi$ to $\psi \phi$ (i.e., $y_{\Psi-\psi} \bar\psi \Psi  \phi$), respectively,
and allowing for the (generic) case of different scales of NP: $M_{\Phi,\Psi,X} \gtrsim \Lambda$ corresponding to
the mass scale of the heavy scalars, vectors and fermions, respectively, we find:
\begin{eqnarray}
f_\Phi\up{2k+4}(u_\Phi,M_\Phi,M) &=& \left| \frac{u_\Phi}{M_\Phi} \right|^2 \left(\frac{-M^2}{M_\Phi^2}\right)^{k} ~, \nonumber \\
f_{\Psi-\psi}\up{2k+4}(y_{\Psi-\psi},M_\Psi,M) &=& \half I_\Psi |y_{\Psi-\psi} |^2 \left(\frac{M^2}{M_\Psi^2}\right)^{k} ~, \nonumber \\
f_X\up{2k+6}(g_X,M_X,M) &=& I_X |g_X|^2  \left(\frac{-M^2}{M_X^2}\right)^{k+1} ~,
\label{eq:fs}
\end{eqnarray}
where $ I_\zeta=1,\,2,\,3$ when the field $\zeta=\Psi~{\rm or}~X$ is an isosinglet, doublet or triplet,
respectively.
Thus, inserting Eq.~\ref{eq:fs} in Eq.~\ref{dmh.eff} and performing the sum over $k$, we obtain
\beq
F\up{\rm eff}(\Lambda)=
\sum_\Phi \frac{|u_\Phi|^2}{M_\Phi^2} A\left( \frac{\Lambda^2}{M_\Phi^2} \right)
+ \half \sum_{\Psi,\psi} I_\Psi |y_{\Psi-\psi}|^2  \left[1-
A\left( \frac{\Lambda^2}{M_\Psi^2}\right) \right]
+ \sum_X I_X |g_X|^2  \left[1-
A\left( \frac{\Lambda^2}{M_X^2}\right) \right] \label{Feffmodels}~,
\eeq
where $A(x)=\ln(1+x)/x$, so that
$1>A(x) \ge0 $, from which it follows that $ F\up{\rm eff} > 0 $.

Assuming that the heavy masses are clustered around a value $M$,
the above expression simplifies to
\beq
F\up{\rm eff}(\Lambda)=  (\xi - \eta) A\left( \frac{\Lambda^2}{M^2} \right) + \eta \,; \qquad
\xi =\sum_\Phi \frac{|u_\Phi|^2}{M_\Phi^2} \,,\quad
\eta= \half \sum_{\Psi,\psi} I_\Psi |y_{\Psi-\psi}|^2
+ \sum_X I_X |g_X|^2  \label{xietaeq}~,
\eeq
where we expect $ \xi,~\eta \sim {\cal O}(1-10)$, e.g., $u_\Phi \sim 3 M_\Phi$
and/or a triplet heavy vector like (colored) quark with $y_{\Psi-\psi} \sim 1$ will give
$\xi,~\eta \sim 10$.

In Fig.~\ref{xieta} we plot the regions in the $\xi - \eta$ plane
which correspond to $\Delta_h = 1$ (natural) and $\Delta_h=10$ (10\% fine-tuning),
for an EFT naturalness scale in the range $3~{\rm TeV} < \Lambda < 10~{\rm TeV}$ and
NP scale $\Lambda < M < 3 \Lambda$. The shaded
region in the $\xi - \eta$ scatter plots correspond to the shaded
regions in the $\Lambda - M$ plane (matching colors). In particular, we
can find the values of ($\xi,\eta$)
for which the EFT corrections to $\Delta m_h^2$ can restore naturalness
in the Higgs sector at a certain $\Lambda$ for some value
$M$ of the NP threshold.
For example, extensions of the SM
with a typical mass scale of $M \sim 7$ TeV that
give $\xi \sim 9$ and $\eta \sim 5$,
will yield an effective
action which is natural up to
$ \Lambda \sim 5$ TeV (an order of magnitude improvement over the pure SM).

\bigskip

\paragraph{Signals of EFT-naturalness} $~$\\
\\
Let us briefly discuss the potential signals of our
EFT-naturalness operators, or equivalently, of the tail of the NP that can
restore naturalness at energy scales which are accessible
to current and future high energy colliders.
In particular, apart from their contribution to
$\delta m_h^2$, these operators also shift the SM Higgs self
couplings $h^3$ and $h^4$, the SM Higgs couplings to the gauge bosons
$hVV$ and $h^2 V^2$ ($V=W$ or $Z$) and the Higgs Yukawa couplings
$h \psi \psi$ ($\psi$ being a SM fermion). In addition,
they also give rise to new higher dimensional
contact terms such as $h^3 V^2$, $h^4 V^2$ and $h^2 \psi^2$.
\begin{table}[htb]
\begin{center}
\begin{tabular}{|c||c|c|c|c|c|c|c|c|c|c|c|c|}
Operator & $h^3$ & $h^4$ & $hWW$ &$h^2W^2$ &$h^3W^2$ & $h^4W^2$ & $hZZ$ & $h^2Z^2$ &$h^3Z^2$ & $h^4Z^2$ & $h \psi \psi$
& $h^2 \psi^2$ \\
\hline \hline
 $\ocal_S\up{2k+4} $& $\checkmark$ & $\checkmark$ & & & & & & & & & & \\
\hline
 $\ocal_{\chibf}\up{2k+4} $& $\checkmark$ & $\checkmark$ & $\checkmark$  & $\checkmark$ & $\checkmark$ &$\checkmark$  & & & & & & \\
\hline
 $\ocal_{\tilde\chibf}\up{2k+4} $& $\checkmark$ & $\checkmark$ & $\checkmark$ & $\checkmark$ & $\checkmark$ & $\checkmark$ & $\checkmark$ & $\checkmark$ & $\checkmark$ &$\checkmark$  & & \\
\hline
 $\ocal_v\up{2k+6}  $&  & &  &  &  & & $\checkmark$ & $\checkmark$ & $\checkmark$ &$\checkmark$  & & \\
\hline
 $\ocal_{\tilde v}\up{2k+6} $&  &  & $\checkmark$ & $\checkmark$ & $\checkmark$ & $\checkmark$ &  &  &  &  & & \\
\hline
 $\ocal_{\VV}\up{2k+6} $&  &  & $\checkmark$ & $\checkmark$ & $\checkmark$ & $\checkmark$ & $\checkmark$ & $\checkmark$ & $\checkmark$ &$\checkmark$  & & \\
\hline
$\ocal\up{2k+4}_{\Psi -\psi}$&  &  &  & & & & & & & & $\checkmark$ & $\checkmark$ \\
\hline
\end{tabular}
\caption{Vertices involving the Higgs, gauge-bosons and fermions which are
generated by the operators in Eqs.~\ref{eq:s-ops}, \ref{eq:v-ops} and \ref{eq:f-ops}.
A check mark is used to indicate that the vertex is affected by the specific operator.}
\label{tab1}
\end{center}
\end{table}

In Table \ref{tab1} we list the expected deviations in the SM couplings
and the new contact terms which are generated
by each of the operators in Eqs.~\ref{eq:s-ops}, \ref{eq:v-ops} and \ref{eq:f-ops}.
Evidently, the tail of the NP generating the EFT-naturalness operators
can be searched for in multi-boson scattering processes
of the form $\psi \bar\psi/VV \to n \cdot h + m \cdot V + X$, where
$n,m=0,1,2,...$. In particular,, one can search for correlations in the various
channels or look for differences between W-boson versus Z-boson
associated production processes. For example, while
$\ocal_{\chibf}\up{2k+4} $ will effect $Wh$ production at the LHC,
i.e., $pp \to Wh + X$, the operator $\ocal_v\up{2k+6}  $
is expected to contribute only to Z-boson associated production processes
such as $pp \to Zh + X$.

Clearly, though, the search
for the tail of these NP effects in Higgs -- gauge-boson processes
will require a sensitivity to these couplings at the percent level,
to be probed at $\Lambda \sim {\cal O}(5-10~{\rm TeV})$.
This will be challenging even at the high-luminosity LHC and may require
future colliders at the high-energy/high-luminosity frontiers, such as
a future 30 or 100 TeV hadron collider and/or an ${\cal O}({\rm TeV})$
$e^+e^-$ collider.
Nonetheless, from our considerations we expect better prospects for detection of NP at the LHC
in the low-multiplicity Higgs -- gauge-bosons production processes, e.g.,
$ pp \to hW,~hZ +X$, and in processes involving the new contact terms
listed in Table \ref{tab1} -  this will be studied in a future work.

\bigskip\paragraph{Constraints from EW precision data and Higgs signals} $~$ \\
\\
Let us now examine the limits that the current data impose on the coefficients
of our
effective operators. Since the most important effects are
generated by the lowest-dimensional operators, we will only investigate the limits
on the dimension 6 coefficients $ f\up6_{\Phi,\Psi-\psi,X}$, which are mainly of two types
[to simplify the expressions we define $ \epsilon = ( v/M)^2$]:

\begin{enumerate}

\item {\bf A shift to the $ \rho $ parameter:}\\
The scalar-triplet operators $\ocal_{\chibf,\tilde \chibf}\up{6}$
and the vector operators $ \ocal\up6_X$
modify the SM gauge-boson masses according to:
\bea
&& \frac{ \delta M_Z^2}{M_Z^2} =  \epsilon \left( f\up6_v + \inv3  f\up6_\VV -  f_{\tilde\chibf}\up6 \right) ~~, ~~
\frac{\delta M_W^2 }{M_W^2}  = \epsilon\left(\half f\up6_{\tilde v}+ \inv3 f\up6_\VV - f_{\chibf}\up{6} - \half f_{\tilde \chibf}\up{6} \right) ~,
\eea
which shift the $\rho $ parameter accordingly:
\bea
\delta \rho = \epsilon\left(\half f\up6_{\tilde v} -f\up6_v +\half f_{\tilde \chibf}\up{6}- f_{\chibf}\up{6}  \right)  ~.
\eea

This is the
strongest constraint from precision EW observables; it is suppressed
by a factor of $ \epsilon $ because the heavy physics being considered here
(the heavy $\Phi$, $X$ and $ \Psi $ states) does not break the SM gauge
invariance (this happens only at the EW scale $v$).
If we assume that the operator coefficients $f$ are $O(1) $ and that there are no cancellations,
then the constraint
$ |\delta\rho| < 0.0007 $ \cite{rhopar-PDG} implies $M \gsim 9.3 $ TeV;
but this can be considerably reduced if some cancellations do occur.

\item {\bf A shift of the Higgs boson couplings to fermions and SM gauge-bosons:}\\
This effect can be divided into three:
\begin{itemize}
\item The scalar operators $\ocal_\Phi\up{6}$ modify the Higgs kinetic term.
Thus, in order
to recover a canonically normalized Higgs field we need to rescale
$ h \to [1 + (\epsilon/2)  \sum f_\Phi\up 6] h $.
This modifies the Higgs couplings to all other SM fields (denoted by $x$):
\begin{equation}
\delta_\Phi \equiv \frac{\delta g_{h xx}}{g_{h xx}^{SM}} =  \frac{\epsilon}{2} \sum_\Phi f_\Phi\up 6 \label{dphi} ~,
\end{equation}
and changes all Higgs decay widths into any final state $x$ by the same factor (so that branching ratios remain the same):
$ \delta \Gamma( h \to xx) \approx \epsilon \sum f_\Phi\up 6 \Gamma_{SM}( h \to xx) $
(to lowest order in $ \Lambda $).
\item The fermion operators $ \ocal_{\Psi-\psi}\up6 $
modify the Higgs Yukawa coupling to the SM fermions ($\psi$):
\begin{equation}
\delta_{\Psi-\psi} \equiv \frac{\delta g_{h \psi \psi}}{g_{h \psi \psi}^{SM}} = -  \epsilon
\sum_\Psi f_{\Psi-\psi}\up 6  \label{dpsi} ~.
\end{equation}
Note that this shift in the $h tt$ coupling also modifies the top-quark loop
contribution in the gluon-fusion Higgs production cross section as well as in the
1-loop decays $h \to \gamma \gamma$ and $h \to Z \gamma$.
\item The scalar-triplet operators $\ocal_{\chibf,\tilde \chibf}\up{6}$ and
the vector operators $ \ocal\up6_X$ modify the couplings of the Higgs to the vector bosons:
\begin{eqnarray}
\delta_W \equiv \frac{\delta g_{hWW}}{g_{h WW}^{SM}} &=&  \epsilon \left( f\up6_{\tilde v} + \frac23 f\up6_\VV  - \frac{1}{2} f\up6_{\chibf} - \frac{1}{4} f\up6_{\tilde\chibf} \right) \label{dW} ~, \\
\delta_Z \equiv \frac{\delta g_{hZZ}}{g_{h ZZ}^{SM}} &=&  \epsilon\left( 2f\up6_{v} + \frac23 f\up6_\VV - \frac{1}{2} f\up6_{\tilde\chibf} \right) \label{dZ} ~,
\end{eqnarray}
and therefore changes the Higgs decay width to a pair of SM gauge-bosons.
\end{itemize}

\end{enumerate}

Turning now to the overall effect of the above modifications on the Higgs couplings to the SM fermions and gauge-bosons,
let us analyze the constraints that can be imposed from the recently measured Higgs signals (see also
\cite{1308.2255}). In particular, defining:
\begin{eqnarray}
w_{xx} \equiv \frac{\Gamma(h \to xx)}{\Gamma(h_{SM} \to xx)} ~,~
R^{Total} \equiv \frac{\Gamma_h^{Total}}{\Gamma_{h_{SM}}^{Total}} \label{ratios} ~,
\end{eqnarray}
the normalized branching ratios for each channel are given by:
\begin{eqnarray}
R^{BR}_{xx} \equiv \frac{BR(h \to xx)}{BR(h_{SM} \to xx)} = \frac{w_{xx}}{R^{Total}} ~.
\end{eqnarray}

The ``signal strength" for each Higgs production and decay mode is then given by:
\begin{eqnarray}
\mu_{xx}^{ii} \equiv \frac{\sigma(ii \to h \to xx)}{\sigma(ii \to h_{SM} \to xx)} =
\frac{w_{ii}  w_{xx}}{R^{Total}} \label{sigst} ~,
\end{eqnarray}
so that:
{\small
\beq
\begin{array}{c||c|c|c|c|c|c|c|c}
{\bf Higgs~signal} & gg \to \gamma \gamma & gg \to Z Z^\star &
gg \to W W^\star & gg \to Z \gamma & gg \to \tau^+ \tau^- &
VV \to \gamma \gamma & q \bar q \to Vh \to V b \bar b &
q \bar q \to V_1 h \to V_1 V_2 V_2^\star   \\
\hline
\boldmath{\mu_{xx}^{ii} \cdot R^{Total} } & w_{gg} w_{\gamma \gamma}  & w_{gg} w_{ZZ}  & w_{gg} w_{WW}  &
w_{gg} w_{Z \gamma}  & w_{gg} w_{\tau \tau} & w_{VV} w_{\gamma \gamma}  & w_{VV} w_{bb}  &
w_{V_1V_1} w_{V_2V_2}
\end{array}
\eeq
}

Note that, since the 1-loop $hgg$ coupling is controlled primarily by the top-quark Yukawa
coupling, we have $w_{gg}=w_{tt}$. To lowest order in $\epsilon$, we then find:
{\small
\beq
\begin{array}{llllll}
w_{\psi \psi}  \approx 1 + 2 \tilde\delta_\psi ~,
& w_{gg}  \approx  w_{tt} ~, & w_{WW}  \approx  1+ 2 \tilde\delta_W ~,
w_{ZZ}  \approx  1+ 2 \tilde\delta_Z ~,
& w_{\gamma \gamma}  \approx   1+ 2.56 \tilde\delta_W - 0.56 \tilde\delta_t ~, &
w_{Z \gamma}  \approx   1+ 2.1 \tilde\delta_Z - 0.1 \tilde\delta_t ~,
\label{omega1}
\end{array}
\eeq
}
and
\begin{eqnarray}
R^{Total}  \approx  1+ 2 \left( {\rm BR}_{SM}^{WW} \, \tilde\delta_W +
{\rm BR}_{SM}^{ZZ}\,  \tilde\delta_Z +
{\rm BR}_{SM}^{gg} \,\tilde\delta_t+
\sum_\psi  {\rm BR}_{SM}^{\psi \psi}  \, \tilde\delta_\psi \right)
  \label{omega2} ~,
\end{eqnarray}
where ${\rm BR}_{SM}^{xx} \equiv {\rm BR}(h_{SM} \to xx )$ are the SM branching ratios,
$\tilde \delta_\psi \equiv \delta_\Phi + \delta_{\Psi-\psi}$,
$\tilde \delta_W \equiv  \delta_\Phi + \delta_W$,
$\tilde \delta_Z  \equiv  \delta_\Phi + \delta_Z$
and $\delta_{\Phi},~\delta_{\Psi-\psi},~\delta_W,~\delta_Z$ are given in Eqs.~\ref{dphi}-\ref{dZ}.

We see that, if the coefficients of the higher dimensional operators
$f_\Psi\up 6, ~ f_\Phi\up 6$ and $f_X\up 6$ are of $ O(1)$, then the typical
correction to the Higgs signal strengths is of $O(\epsilon)$.
Thus, given that the LHC is expected to probe the Higgs couplings to at most
10\% accuracy (this includes the high luminosity run of the LHC \cite{1403.7191}),
a rather weak bound of $M \gtrsim O(1~{\rm TeV})$,
can be imposed on the scale of the new heavy physics that can
lead to EFT-naturalness in the Higgs sector.
A future $e^+e^-$ collider may be able to improve the accuracy
to $O(0.01)$ \cite{1403.7191}, in which case
EFT-naturalness can be probed up to $\Lambda \sim 2-3$ TeV.
We conclude that precision Higgs measurements are not expected
to impose significant constraints on our EFT-naturalness scenario.\\

Finally, we note that none of the operators in Eqs.~\ref{eq:s-ops}, \ref{eq:v-ops} and \ref{eq:f-ops}
breaks any of the global symmetries of the SM, so no strong limits can be obtained
from CP, flavor or lepton and baryon number conservation experiments.

\bigskip\bigskip\paragraph{Summary} $~$\\
\\

We have used EFT techniques to study
the little hierarchy problem of the SM Higgs sector
assuming the presence of weakly-coupled and decoupling heavy physics with scale $M$.
Following Wilson's approach we determine the conditions under which
the quadratic heavy-physics contributions can balance those generated by the SM
at the scale of the effective action $ \Lambda $ ($\Lambda \gg m_W$).

We analyze the complete set of higher dimensional effective operators
(at any dimension $n \ge 5$) that can yield ${\cal O}(\Lambda^2)$ contributions
to $\delta m_h^2$ in the EFT and
classify the underlying heavy theories that can generate these operators at tree-level.
In particular, we find that heavy new physics theories that can lead to EFT-naturalness
(i.e., that can restore naturalness in the effective action
at e.g., $\Lambda \sim 5 -10$ TeV)
must contain one or more singlet or triplet heavy
bosons or else a singlet, doublet or triplet fermions, all heaving typical masses larger than $\Lambda$.
We then estimate the coefficients of the EFT-naturalness higher dimensional operators
using the relevant phenomenological interactions of these heavy particles.

We have also studied the constraints that precision electroweak data and the recently measured
Higgs signals impose on our EFT-naturalness setup and find that
heavy scalar singlets and/or heavy fermions (singlets, doublets or triplets)
are more likely to play a role in softening the fine-tuning in the SM Higgs sector,
if the scale of the
new heavy physics is below $\sim 10$ TeV.

Finally, we have discussed the expected signatures that the tail of the NP (responsible for EFT-naturalness)
can have at the LHC and at future colliders. In particular,
we find that signals of EFT-naturalness are likely to be manifest
as deviations in processes involving Higgs + gauge-boson production,
e.g., $pp ~ {\rm or} ~ e^+e^- \to hh,hW,hZ,WW,ZZ + X$ and/or
processes with higher
Higgs/gauge-boson multiplicities in the final state.

\bigskip

{\bf Acknowledgments:}
We thank Michael Trott for making useful comments on the manuscript.
The work of AS was supported in part by DOE contract
\#DE-AC02-98CH10886(BNL) and
JW is supported in part by a UCR CoR grant.


\end{document}